\documentclass{aa}
\usepackage{graphicx}
\usepackage{natbib}
\bibpunct{(}{)}{;}{a}{}{,}

\newcommand{\brg}{Br$\gamma$}

\newcommand{\pb}{P$\beta$}

\newcommand{\hei}{He\,{\sc i}}

\usepackage{amssymb}
\newcommand{\mum}{$\mu$m}

\begin{document}
%
%\headnote{Letter to the Editor}

   \title{Identification of the ionizing source of NGC
2024\thanks{Based on observations collected at the European Southern
Observatory at La Silla and Paranal, Chile (ESO programmes 62.H-0443
and 64.H-0425)}}

   \author{A.~Bik\inst{1} \and A.~Lenorzer\inst{1} \and
   L.~Kaper\inst{1} \and F.~Comer\'on \inst{2} \and L.B.F.M.~Waters
   \inst{1,3} \and A.~de Koter \inst{1} \and M.M.~Hanson\inst{4} }

   \offprints{A. Bik (bik@science.uva.nl)}

   \institute{Astronomical Institute ``Anton Pannekoek'',
           University of Amsterdam, \\ Kruislaan 403, 1098 SJ Amsterdam,
           The Netherlands
         \and European Southern Observatory, Karl-Schwarzschild
           Strasse 2, Garching-bei-M\"unchen, D85748, Germany
         \and Instituut voor Sterrenkunde, Katholieke Universiteit Leuven,
           Celestijnenlaan 200B, B-3001 Heverlee, Belgium
         \and University of Cincinnati, Cincinnati, OH 45221-0011, U.S.A.
             }

   \date{Received; accepted}

\authorrunning{A.\ Bik et al.}
\titlerunning{The ionizing source of NGC~2024}

\abstract{We propose the late-O, early-B star IRS2b as the ionizing
source of the Flame Nebula (\object{NGC~2024}). It
has been clear that such a hot, massive star must be present in this
heavily obscured region, and now it has been identified. New
near-infrared photometry shows that IRS2b is the most luminous and
hottest star in the young star cluster embedded in the center of
NGC~2024. The near-infrared observations ($5' \times 5'$) cover
$\sim$90~\% of the H~{\sc ii} region detected in radio continuum
radiation, making the probability very low that the ionizing star is
not present in the field. A K-band spectrum of IRS2b obtained with
ISAAC on the {\it Very Large Telescope} indicates that the spectral
type of IRS2b is in the range O8~V -- B2~V. Additional arguments
indicate that its spectral type is likely closer to O8 than to B2.
 The corresponding amount of ionizing radiation is consistent with
 published radio continuum and recombination line observations.  
}

   \maketitle
%
%________________________________________________________________

\section{Introduction}

Just east of \object{$\zeta$~Ori} and north-east of the Horsehead
Nebula, \object{NGC~2024} (Flame Nebula) 
appears in visible light as a bright nebula of which the
central part is obscured by a thick dust lane. The central region of
\object{NGC~2024} is also a bright source of radio continuum emission
and recombination lines \citep{Krugel82,Barnes89}, indicating the
presence of an ionizing star of spectral type O9--O9.5. The heavy
obscuration by dust is the reason why, contrary to \object{M42}
\citep{Odell01} and many other visible H~{\sc ii} regions, the
exciting star(s) of \object{NGC~2024} have not been identified. Near-
and mid-infrared observations have revealed the presence of a young
star cluster in the core of NGC~2024
\citep{Barnes89,Lada91,Comeron96Flame,Haisch01}. Several candidates
for the ionizing source of \object{NGC~2024} have been proposed
\citep{Grasdalen74,Barnes89}, but none of these candidates 
are able to produce the observed radio continuum emission.

We are carrying out a systematic survey of the stellar content of
compact and ultra-compact H\,{\sc ii} regions, with the aim to detect
and study deeply embedded, newly-born massive stars. Our ultimate
goal is to better understand the earliest phases in the life of the
most massive stars. In this context we have obtained deep
near-infrared images of a sample of compact H~{\sc ii} regions,
including \object{NGC~2024} \citep{Kaper03}. Subsequent K-band
spectroscopy of the candidate ionizing stars has resulted in the determination of their spectral types
\citep{Ostarspec03}.

Here we describe how we have identified the long-sought-for ionizing
star of \object{NGC~2024}. In Sect.~\ref{sec:obs} we present
near-infrared, narrow-band images of this region, as well as a K-band
spectrum of the candidate ionizing star IRS2b. In Sect.~\ref{sec:prop}
we derive the physical properties of this star. In the last section we
present our conclusions and compare the derived properties of
IRS2b with alternative, previously proposed candidates.

\section{Observations}\label{sec:obs}

Near-infrared images of the central region of \object{NGC~2024} were
obtained with SOFI mounted on ESO's \emph{New Technology Telescope}
(NTT) at La Silla on February 6, 1999; the seeing was $0.7''$. We used
narrow-band filters centered on strong nebular emission lines
(\pb\ 1.28~\mum, H$_2$ 2.12~\mum, Br$\gamma$ 2.16~\mum), and two
narrow-band continuum filters in the J and K-band (1.21 and
2.09~\mum). The latter were chosen to measure the J and K magnitude of 
the embedded stars, avoiding the contamination by nebular emission 
lines.

Nine frames of 20 seconds each were taken on source and 9 frames with a 2
second exposure per frame were obtained to measure the sky
background. 
 The observations were reduced using standard infrared imaging data
 reduction procedures with IRAF. 

Pointsources were detected by adding the frames in the J and K
continuum filters, and running DAOPHOT \citep{Stetson87} on the
resulting frame. The photometry was obtained as follows: 
aperture photometry with a large aperture was performed on the image
of the standard star and of bright, isolated stars in the image
field. This allowed us to set up a network of bright secondary
standards in the image field. Then, aperture photometry with a small
aperture (3 pixel in radius), adequate for our rather crowded field,
was performed on all the stars in the field. The magnitudes of these
stars were determined taking those of the secondary standards as a
reference.

We note that for red 
sources ((J-K)~$\gtrsim 2$~mag) the narrow-band J and K magnitudes 
differ significantly from the broad-band J and Ks magnitudes. This 
difference has been estimated by multiplying various energy
distributions with the response curves of the  
filters, representing a range in (J-K) from 0 to 7. The typical
errors for (J-K) $\approx 5$ on this
correction are 0.2 and 0.07 mag in J and K, respectively.
The J and K-band magnitudes used in this paper are based on narrow-band
observations. Only for the construction of the 
Hertzsprung-Russell diagram (HRD, Fig.~\ref{fig:hrd}) the narrow-band 
magnitudes are converted into broad-band magnitudes.

Medium-resolution ($R \simeq 8,000$) K-band spectra of some stars in
this region were taken with ISAAC and ESO's \emph{Very Large
Telescope} (VLT) at Paranal, Chile on March 20, 2000.  The spectra
were dark- and flat-field corrected, and wavelength calibrated using
standard reduction procedures. In order to correct for the sky
background the object was ``nodded'' between two positions on the slit
(A and B) such that the background emission registered at position B
(when the source is at position A) is subtracted from the source plus
sky background observations at position B in the next frame, and vice
versa. Telluric absorption lines were removed using the telluric
standard star \object{HD 39908} with spectral type A2~V observed under
identical sky conditions.  
The only photospheric line (\brg) in the spectrum of the telluric star
  needs to be devided out first. It turns out that the best result is
  achieved when first the telluric features are removed from the
  K-band spectrum of the   telluric standard using a high resolution
  telluric spectrum (obtained by NSO/Kitt Peak). This spectrum is taken
  under completely different sky conditions, so a lot of remnants are
  still visible in the corrected standard star spectrum; without this
  ``first-order'' telluric correction, a proper fit of \brg\ cannot be obtained.
 The \brg\ line is fitted by a
combination of two exponential functions. The resulting error on the
\brg\ equivalent width (EW) of our target star is about 5\%.

 An L-band spectrum of IRS2b, with a spectral resolution
 $R \simeq 1\,200$ was obtained with VLT/ISAAC on February 22, 2002. 
 The data reduction was performed using an A6\,II star (HD\,73634) 
 as telluric standard. The quality of the telluric standard was not
 sufficient to obtain a quantitative measurement of the hydrogen
 absorption lines in the L band  spectrum.

\section{The nature of the ionizing source in NGC~2024}\label{sec:prop}

Fig.~\ref{fig:color_image} shows a composite near-infrared image of
the obscured central region of \object{NGC~2024} ($5' \times 5'$),
centered on the bright (saturated, K~$\sim 5$) infrared source IRS2. 
Exposures in three narrow-band filters are
combined: \brg, H$_2$ and \pb. The edge of the
molecular cloud to the south of NGC~2024 is marked by emission
produced by excited molecular hydrogen: note the clumpy filamentary
structure. The recombination lines of hydrogen (e.g.\ \brg) are strongest towards
the core of the embedded stellar cluster. The optical extent of the
Flame Nebula is roughly $15' \times 15'$, but at radio wavelengths NGC
2024 is only slightly larger ($6' \times 6'$) than the field covered
by our near-infrared observations. The radio continuum radiation is
strongly concentrated towards the central part of NGC~2024, and is
expected to outline the extent of the region where hydrogen is fully
ionized. 

In the following we demonstrate that IRS2b is the  best candidate
  ionizing star  present
in the central star cluster of NGC~2024. A K-band spectrum of IRS2b is
used to confirm its early spectral type. The resulting effective
temperature and luminosity of IRS2b are consistent with the amount of
observed radio continuum and line emission.

\begin{figure*}
%Fig. 1
%\centering \resizebox{\hsize}{!}{\includegraphics{image.eps}}
   \caption{Composite of three narrow-band images of the central
   region of \object{NGC~2024} obtained with NTT/SOFI (blue: \pb; red:
   \brg; green: H$_{2}$). North is up and East to the left; the image
   size is $5' \times 5'$ (at the distance of NGC2024 this corresponds
   to 0.5 $\times$ 0.5 pc). The upper left panel zooms in on the region
   around IRS2: about $5''$ to the north-west we find IRS2b, the
   ionizing star of \object{NGC~2024}. Also some previously proposed
   candidate ionizing stars are indicated. The vertical stripes are
   instrumental artifacts caused by the brightest stars.}
   \label{fig:color_image}
\end{figure*}
\subsection{The young stellar cluster embedded in NGC~2024}

Due to the severe obscuration by the dust lane in front of NGC~2024,
its stellar population becomes apparent at near-infrared wavelengths. 
The strong infrared sources IRS1-5 are the brightest objects in
Fig.~\ref{fig:color_image}.  Another bright source, IRS2b is located
just  $5''$ north-west of IRS2 (see inset Fig.~\ref{fig:color_image}).  
The existence of IRS2b was first reported by \citet{Jiang84} and
confirmed by \citet{Nisini94}.

 We detect 106  and 210
point sources down to the detection limit of 17.7~mag in J and of
16.1~mag in K, respectively. Because of the proximity of NGC~2024 
 \citep[360 pc,][]{Brown94}, its
small spatial extent, its angular separation from the galactic plane,
and the obscuration of background sources by the molecular cloud, we
expect that nearly all of the infrared sources present in our
near-infrared image are physically related to the cluster 
\citep[cf.][]{Comeron96Flame}. 

The brightest infrared source in the center of the NGC~2024 cluster is
IRS2, a source that has been held responsible for the ionization of
the H~{\sc ii} region since its discovery
\citep{Grasdalen74}. However, the physical nature of IRS2 is unclear:
its K-band magnitude is too bright for a main sequence star,
indicating a strong infrared excess. This excess is likely due to the
presence of a dense circumstellar disk; the central star is probably a
B-type star \citep[][ and below]{Lenorzer03}. Can we identify
another, potentially hotter star that could be responsible for the ionization of
the Flame Nebula? 

\begin{figure}
%Fig. 2
   \centering
   \resizebox{\hsize}{!}{\includegraphics{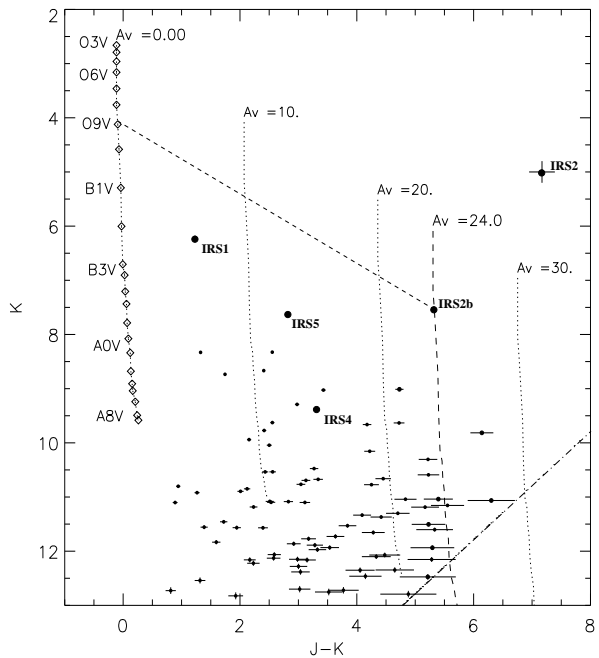}}
   \caption{Color-magnitude diagram of the brightest stars in the
   central region of \object{NGC~2024}. The vertical (dotted) lines
   indicate the position of the main sequence for different amounts of
   visual extinction (A$_{V}$). The magnitudes are obtained through
   narrow-band observations. The diagonal (dashed) line is the
   dereddening line of IRS2b, demonstrating that IRS2b is the
   intrinsically brightest star of the embedded population. IRS2
   (saturated in K) probably is a B star with a strong infrared excess
   due to a circumstellar disk \citep{Lenorzer03}. The
   dashed-dotted line is the detection limit in (J-K) following from the
   detection limits in J and K.  } \label{fig:cmd}
\end{figure}

The K vs. (J-K) color-magnitude diagram (CMD) of the stars detected in
our images is presented in Fig.~\ref{fig:cmd}; only the brightest
objects are shown.  The dotted line indicates the position of the main
sequence \citep{LandoltBornstein}, for different amounts of visual
extinction A$_V$. A distance modulus of $7.80\pm 0.45$~mag is adopted,
which corresponds to $363 \pm 75$~pc \citep{Brown94}. An anomalous
extinction law with $R_V = 5.5$ is used, as measured by
\citet{Lee68}. We used the parameterization of
\citet{Cardelli89} to describe the shape of the extinction law.
 This extinction law is derived for stars with
$\mbox{E(B-V)} \leq 1$ and could be different for higher amounts of
reddening. However, the dereddening of OB stars in the near infrared
is not very sensitive to the precise value of $R_V$, as $A_K =
0.108 \times A_V$ for $R_V = 3.1$ and $A_K = 0.125 \times A_V$ for
$R_V = 5.5$. The slope of the near-infrared extinction law changes 
very slowly with $R_V$ and the intrinsic (J-K) color of OB stars is
almost constant with spectral type and luminosity class
\citep{Koornneef83}.

Fig.~\ref{fig:cmd} shows that the star IRS2b (K=$7.57 \pm 0.07$), when dereddened to
the main sequence, is the  best candidate ionizing star in the field (neglecting
IRS2) at a position consistent with a late O main sequence star. 
 The position of IRS2b in the CMD implies $A_V = 24.0\pm 0.5$~mag ($A_V = 28.5 \pm 0.5$
for $R_V = 3.1$) if it is a main sequence star.

\subsection{K-band spectral type of IRS2b}

\begin{figure}[!t]
%Fig 3.
   \resizebox{\hsize}{!}{\includegraphics{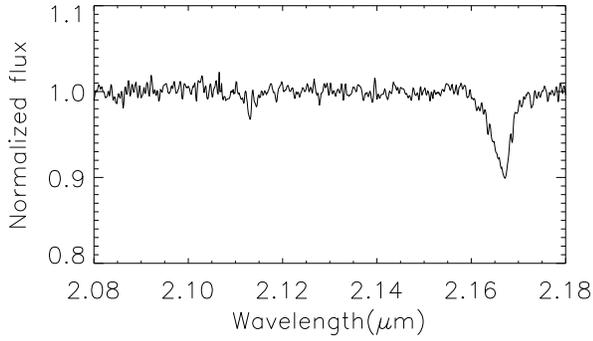}}
   \caption{K-band spectrum of IRS2b, obtained with VLT/ISAAC, showing
   \hei\ 2.1127, 2.1138~\mum\ and \brg\, 2.166~\mum\ in absorption.}
   \label{fig:spectrum}
\end{figure}

The spectral type of IRS2b can be determined from its K-band spectrum
(Fig.~\ref{fig:spectrum}).  \citet{Hanson96} define five K-band
spectral classes for O and early-B stars based on low-resolution
K-band spectra. Two K-band spectral classes correspond to the hottest
O stars (kO3--O4 and kO5--O6, with k denoting that the classification
is based on the K-band spectrum) which show lines of N\,{\sc iii}
(2.115~\mum) and C\,{\sc iv} (2.079~\mum) in emission and the \brg\
(2.166~\mum) line in absorption. In the third class (kO7-O8) the
C\,{\sc iv} line is absent and the \hei\ (2.1128, 21137~\mum) lines
appear in absorption.  The fourth spectral class kO9--B1 is defined as
having \hei\ and \brg\ both in absorption, with the equivalent width
of \brg\ less than about 4~\AA. This class is equivalent to a
Morgan-Keenan (MK) spectral type between O8~V and B1~V. The fifth class
(kB2--B3) shows strong \brg\ and \hei\ absorption lines and is
equivalent to MK type B1~V to B2~V. The line equivalent widths (EW)
can in principle be used to determine the K-band spectral type. There
is, however, a substantial scatter in the observed EW of \brg\ as a
function of spectral type in the calibration stars used by
\citet{Hanson96}.

IRS2b has both \hei\ 2.113~\mum\ and \brg\ in absorption, with an
equivalent width of $0.4 \pm 0.1$~\AA\ and $4.9 \pm 0.6$~\AA,
respectively. The \brg\ EW indicates that the K-band spectral type of
IRS2b is in the range kO9--B1 to kB2--B3 (MK types O8--B2). The \hei\
line appears in absorption around spectral type O7.5, increasing its
EW towards later spectral types. The \hei\ line decreases in strength
again towards B2/B3. The relatively weak \hei\ line in
the spectrum of IRS2b suggests that its spectral type is either close
to O8, or more towards B2/B3.

An L-band spectrum provides a better diagnostic of its spectral type,
since it includes hydrogen Pfund lines, which are more temperature
sensitive than \brg\ \citep{Lenorzer02Atlas}. The L-band spectrum we
obtained from IRS2b is of insufficient quality to measure the strength
of the absorption lines, though it is clear that they are
present.  Although we are not able to provide an independent
  estimate of its spectral type, the L-band spectrum rules out the
  possibility that IRS2b has an infrared excess. Such an excess would
  be even more dominant in the L-band, hindering the detection of photospheric lines.

\subsection{The position of IRS2b in the HRD}

Taking our measurement errors into account, we arrive at a
(conservative) estimate of the MK spectral type of IRS2b in the range
between O8~V and B2~V. This corresponds to a range in $T_{\rm eff}$
between 34,000 and 22,000~K \citep{Martins02,LandoltBornstein}. 
  Note, however, that the effective temperature calibration of OB
  stars is still a matter of debate. The derived limits on 
$T_{\rm eff}$  are not very hard.

\begin{figure*}
%Fig. 4.
\centering \resizebox{\hsize}{!}{\includegraphics{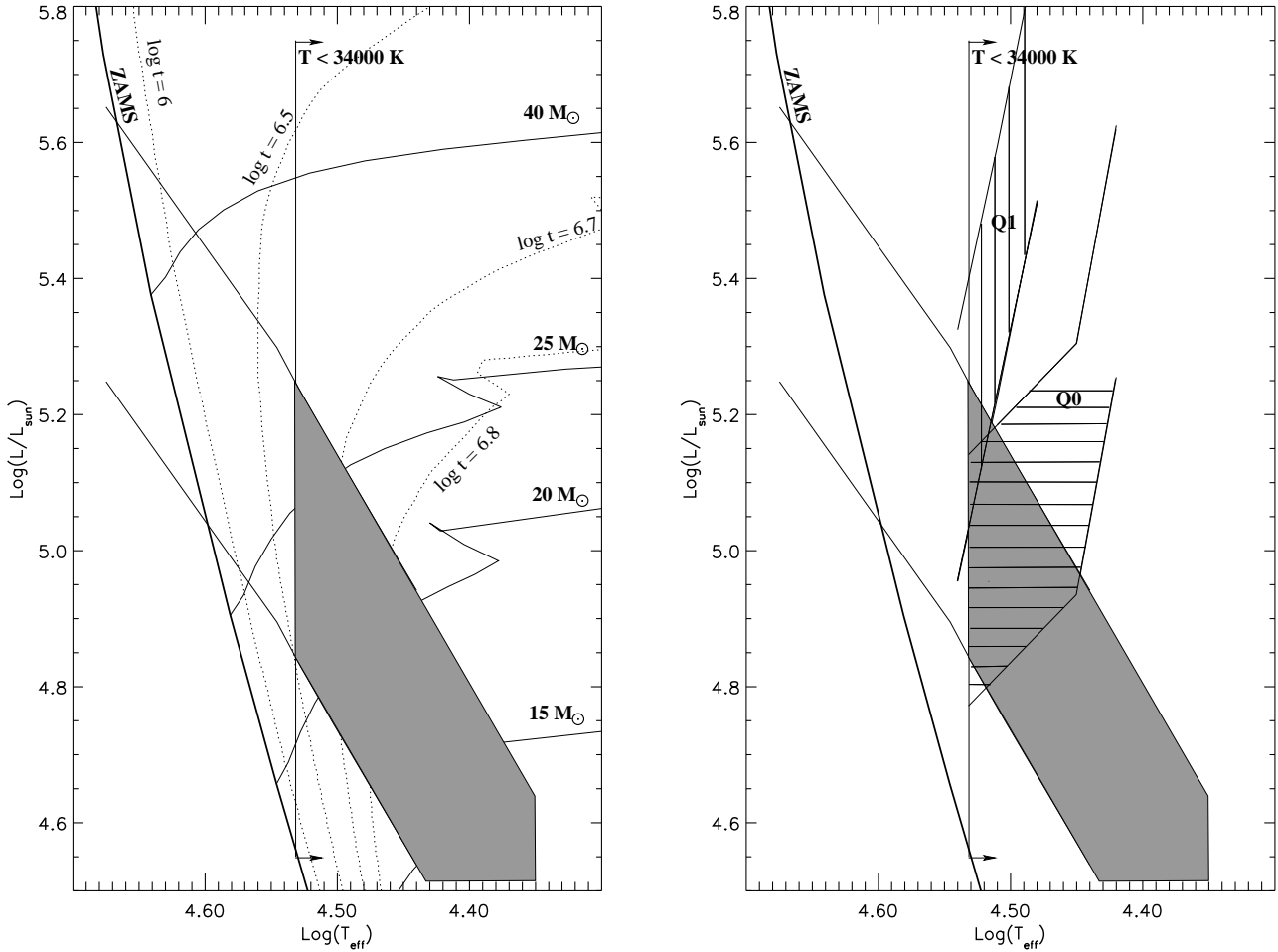}}
 \caption{\emph{Left:}The upper part of the theoretical HRD. The shaded region
 indicates the position of IRS2b for which all observational
 constraints on $L$ and $T_{\rm eff}$ are fulfilled. The diagonal 
area represents the constraint provided by the observed K-band
 magnitude. The K-band spectrum implies that $T_{\rm eff} \leq 34000$
 K.  Also shown are the theoretical evolutionary tracks and isochrones from \citet{Lejeune01}.
 \emph{Right:} The same region of the HRD  as shown in the left
 panel, but now including the regions  in the HRD capable to produce Q$_0$ and
 Q$_1$ for comparison.}
\label{fig:hrd}
\end{figure*}

For every $T_{\rm eff}$, the dereddened K-band magnitude can be
converted into a luminosity (adopting a distance of 363 pc).  For this
calculation we use the relation between the bolometric correction and
$T_{\rm eff}$ from \citet{Vacca96}  and the (V-K) vs. 
$T_{\rm eff}$ relation given by \citet{Koornneef83}.  The uncertainty in the luminosity
of IRS2b is due to the uncertainty in the distance towards
\object{NGC~2024} and to the conversion from narrow-band to broad-band
magnitudes. The observational constraints on the location of IRS2b in
the  HRD are visualized in Fig.~\ref{fig:hrd} (left panel, shaded area).

%ew of HeI line: 0.38+- 0.13
%ew of HI line: 4.9 +- 0.6

Are these constraints on $L$ and $T_{\rm eff}$ of IRS2b consistent
with the observed nebular emission?  The radio continuum flux of the
H~{\sc ii} region can be used to estimate the ultraviolet flux of the
ionizing source of \object{NGC~2024}.  The total radio continuum flux
at 1667~MHz derived from high-resolution radio observations
\citep{Barnes89} is $63 \pm 4$~Jy. The number of Lyman continuum photons is
derived from the observed radio continuum flux under the assumption
that the H~{\sc ii} region is ionization bounded and neglecting
absorption of UV radiation by dust, implying that the number of
recombinations is equal to the number of ionizations. We find that the
number of Lyman continuum photons is 
$7.3 \pm 1.2 \times 10^{47}$  The given accuracies reflect the error
in the distance quoted by \citet{Brown94}.

\citet{Krugel82} performed observations of the radio recombination
lines H76$\alpha$ and He76$\alpha$. The corresponding ionized helium
fraction is 3~\%, which leads to $2.2 \pm 0.1 \times 10^{46}$ helium
continuum photons.
 
The observed number of photons capable of ionizing hydrogen (Q$_0$)
and helium (Q$_1$) can be compared to predictions based on stellar
atmosphere models of OB stars \citep{Smith02}. If we assume that a
single hot star ionizes \object{NGC~2024}, Q$_0$ and Q$_1$ constrain
the location of the ionizing star in the HRD. In Fig. \ref{fig:hrd}
(right panel), the regions which are able to produce the required
amount of Q$_0$ and Q$_1$ are indicated by vertical and horizontal
stripes, respectively. These regions overlap with the shaded 
region in the HRD defined by the spectral classification of IRS2b.

\section{Discussion}\label{sec:conc}

As shown in the previous section, an O8 star with characteristics
compatible with those observed in IRS2b should be  capable of
producing the required amount of ionizing radiation to explain the
degree of ionization of NGC~2024. To meet the constraint set by Q$_1$,
its spectral type is more likely to be late O than early B, which is
supported by its position in the CMD. Note, however, that the predicted
value for Q$_1$ is not well known and depends on the selected stellar model.
 Although other candidate ionizing stars exist, they are not needed to
 provide a self-consistent solution for the ionization of NGC~2024. In
 the following we will discuss some of these alternatives. 

If we correct for foreground extinction (cf.\ Fig. \ref{fig:cmd}),
IRS2b and IRS2 are the two brightest sources.  The position of IRS2 in
the CMD indicates a strong infrared excess
\citep{Lenorzer03,Grasdalen74}. IRS2 is associated with the ultra-compact
radio source \object{G206.543-16.347}, but its physical nature is
unclear. Current models suggest that ISR2 is a B star surrounded by a
dense circumstellar disk responsible for the production of the
infrared excess.

IRS1 \citep[estimated spectral type B0.5V,][]{Garrison68} and IRS2 have been
candidate ionizing sources for a long time.  \citet{Barnes89}
discovered 29 additional near-infrared sources in the Flame Nebula. In
Figs.~\ref{fig:color_image} and \ref{fig:cmd}, the brightest of these
infrared sources discovered by \citet{Barnes89} are plotted, as well
as IRS1 and IRS2. They propose that IRS1, IRS4 and IRS5, together with
IRS3, provide a significant contribution to the ionizing
radiation. IRS3 is not shown in the CMD, because it is not a single
source, but consists of multiple stars which were not resolved by
\citet{Barnes89}. As $Q_1$ is decreasing very rapidly with later
  spectral type the contribution to $Q_1$ by the other bright stars 
(all later spectraltype than B0.5V) is 
  negligible.

The radio continuum emission, however, can only be transformed into the
number of ionizing photons assuming that there is no dust included in
the H\,{\sc ii} region, and that the H\,{\sc ii} region is ionization
bounded. If these assumptions are not valid, the derived values of
Q$_0$ and Q$_1$ are lower limits, so that a star hotter than IRS2b
might be required. In principle, such a star could be significantly
more reddened (A$_V~\ge~35$ mag, based on our detection limit in J)
and thus have remained undetected, especially when the amount of
extinction strongly varies with position. However, the visual
extinction we measure for the stars in the field is in the range
between $5.5 \leq A_V \leq 25$~mag. We judge that the probability
of selectively obscuring this potentially hotter star is low.

In Fig. \ref{fig:hrd} evolutionary tracks and isochrones from
\citet{Lejeune01} are plotted and the zero-age main sequence (ZAMS) is
indicated. The location of IRS2b suggests that its mass is between 15
and 25 M$_{\sun}$. Although it is not possible to determine the
  age of a single star, to position of IRS2b in \ref{fig:hrd} is,
  given the uncertainty in the effective temperature calibration
  consistent with the age proposed
by \citet{Blaauw91, Brown94} and \citet{Comeron96Flame}.  
 \citet{Comeron96Flame} derive an age of $2 \times 10^6$ years
  based on the frequency of IR excess in very low mass objects in NGC~2024 as
  compared to that in Rho Ophiuchi.

\begin{acknowledgements}
AB acknowledges financial support from the DFG during a two-month
visit at ESO Headquaters. LK is supported by a fellowship of the Royal
Academy of Arts and Sciences in the Netherlands. We would like to thank
L. Decin for her help in the data reduction of the L band
spectrum. We thank the anonymous referee for critical and constructive
comments. NSO/Kitt Peak FTS data used here were produced by NSF/NOAO.
\end{acknowledgements}

%\bibliographystyle{apj}
%\bibliography{/home/bik/Papers/arjan}

\end{document}